\documentclass[twocolumn,aps,floatfix]{revtex4}
\usepackage{latexsym,amsmath,amssymb,amsbsy}
\usepackage{graphicx}
\usepackage{nccfoots}
\usepackage[symbol]{footmisc}

\begin{document}


\title{Dissociation of $^{10}$C Nuclei in a Track Nuclear Emulsion \\
at an Energy of 1.2 GeV per Nucleon}


\author{\textbf{K.~Z.~Mamatkulov$^{\textbf{1), 2)}}$, R.~R.~Kattabekov$^{\textbf{1), 3)}}$, S.~S.~Alikulov$^{\textbf{2)}}$, D.~A.~Artemenkov$^{\textbf{1)}}$, R.~N.~Bekmirzaev$^{\textbf{2)}}$, V.~Bradnova$^{\textbf{1)}}$, P.~I.~Zarubin$^{\textbf{1)~*}}$, I.~G.~Zarubina$^{\textbf{1)}}$, N.~V.~Kondratieva$^{\textbf{1)}}$, D.~O.~Krivenkov$^{\textbf{1)}}$, A.~I.~Malakhov$^{\textbf{1)}}$, K.~Olimov$^{\textbf{1), 3)}}$, N.~G.~Peresadko$^{\textbf{4)}}$, N.~G.~Polukhina$^{\textbf{4)}}$, V.~V.~Rusakova$^{\textbf{1)}}$, R.~Stanoeva$^{\textbf{1), 5)}}$, and S.~P.~Kharlamov$^{\textbf{4)}}$}
\Footnotetext{1)}{Joint Insitute for Nuclear Research, Dubna, Moscow Region, 141980 Russia}
\Footnotetext{2)}{A. Kodirii Jizzakh State Pedagogical Institute, st. S.Rashidov 4, Jizzakh, 130114 Republic of Uzbekistan}
\Footnotetext{3)}{Institute for Physics and Technology, Uzbek Academy of Sciences, ul. G. Mavlyanova 2b, Tashkent, 700084 Republic of Uzbekistan}
\Footnotetext{4)}{Lebedev Physical Institute, Russian Academy of Sciences, Leninskii pr. 53, Moscow, 119991 Russia}
\Footnotetext{5)}{South–West University, Ivan Michailov str. 66, 2700 Blagoevgrad, Bulgaria}
\Footnotetext{*}{E-mail: \texttt{zarubin@lhe.jinr.ru}}}

\indent \par
\noindent \affiliation{Received March 21, 2012}

\begin{abstract} 
\noindent The charge topology in the fragmentation of $^{10}$C nuclei in a track nuclear emulsion at an energy of 1.2 GeV per nucleon is studied. In the coherent dissociation of $^{10}$C nuclei, about 82\% of events are associated with the channel $^{10}$C $\rightarrow$ 2$\alpha +$ 2\emph{p}. The angular distributions and correlations of product fragments are presented for this channel. It is found that among $^{10}$C $\rightarrow$ 2$\alpha +$ 2\emph{p} events, about 30\% are associated with the process in which dissociation through the ground state of the unstable $^9$Be$_{g.s.}$ nucleus is followed by $^8$Be$_{g.s.} +$ \emph{p} decays.\par

\indent \par
\noindent \textbf{DOI:} 10.1134$/$S1063778813100141
\end{abstract}

\maketitle

The acceleration of stable nuclei and the subsequent separation of their fragmentation and charge-exchange products make it possible to create beams of relativistic radioactive nuclei. At the same time, a track nuclear emulsion irradiated with light relativistic nuclei opens new possibilities for studying these nuclei owing to a complete observation of their fragmentation products. A unification of the above possibilities provides quite an efficient approach to extending the investigation of the structure of nuclei. Conclusions concerning special features of light nuclei are based on the probabilities for observing dissociation channels and on measurements of angular distributions of relativistic fragments. The potential of the spectroscopy of fragmentation final states is determined, first of all, by the accuracy of angular measurements. The angular resolution ensured by the track-emulsion method has a record value not poorer than 10$^{-3}$ rad. The accuracy in measuring fragment momenta is not as critical as the angular resolution, and we assume in our analysis that relativistic fragments conserve the primary-nucleus momentum per nucleon. The data presented in this article on the dissociation of $^{10}$C nuclei by the Becquerel Collaboration \cite{01} permit making the next step in studying the cluster structure of neutron-deficient light nuclei \cite {02,03,04,05,06}.\par
\indent The $^{10}$C nucleus is of particular interest as a source of information about the role of unstable nuclei in the cluster structure. This nucleus is the only example of a stable structure of four clusters that belongs to the super-Boromean type, since the removal of one of the clusters or nucleons leads to an unbound state. The threshold for the reaction $^{10}$C $\rightarrow$ 2$\alpha +$ 2\emph{p}, which leads to the production of an unbound system, is 3.73 MeV. The next dissociation channel is 3.82 MeV and corresponds to the $^8$Be$_{g.s.} +$ 2\emph{p} channel. The knockout of one of the protons (the threshold is 4.01 MeV) leads to the formation of an unstable $^9$Be nucleus, which decays to a proton and a $^8$Be$_{g.s.}$ nucleus. A $^6$Be$_{g.s.}$ resonance may arise with a threshold of 5.10 MeV upon the separation of an alpha-particle cluster from the $^{10}$C nucleus. For this resonance, the decay energy is 1.37 MeV. The decay of the $^6$Be$_{g.s.}$ resonance to the $^5$Li$_{g.s.}$ resonance is impossible since the threshold for the production of the $^5$Li$_{g.s.} +$ \emph{p} system is 0.35 MeV higher than the ground state of $^6$Be. Moreover, the decay $^9$B$_{g.s.} \rightarrow ^5$Li$_{g.s.} + \alpha$, whose threshold is 1.5 MeV higher than the $^9$B$_{g.s.}$ ground state, is impossible because of the smallness of the energy window (185 keV). Therefore, the $^6$Be$_{g.s.}$ and $^5$Li$_{g.s.}$ resonances must arise directly upon the dissociation of $^{10}$C nuclei.\par
\begin{figure}[!ht]
\begin{center}
\includegraphics[width=0.45\textwidth]{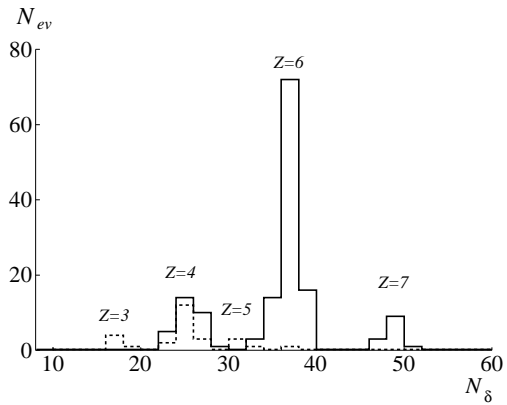}
\caption{Distribution of the numbers of (solid - line histogram) beam - nucleus and (dashed - line histogram) secondary - fragment tracks with respect to the number of delta electrons, \emph{N$_{\delta}$}, per 1 mm of length in events measured in a track emulsion irradiated with a mixed beam of $^7$Be, $^{10}$C, and $^{12}$N nuclei.}
\label{fig:1}
\end{center}
\end{figure}

\indent A stack of layers of BR-2 track nuclear emulsion was irradiated with a mixed beam of $^7$Be, $^{10}$C, and $^{12}$N nuclei that was created by selecting products of charge-exchange and fragmentation processes involving $^{12}$C nuclei accelerated to an energy of 1.2 GeV per nucleon at the nuclotron of the Joint Institute for Nuclear Research (JINR, Dubna)\cite {07}. Interactions where the total charge of relativistic fragments in the respective events satisfied the condition $\Sigma$\emph{Z}$_{fr} >$ 3 were sought in layers of irradiated emulsion via scanning along the tracks of beam nuclei. The charge classification of beam-nucleus and secondary-fragment tracks in the events under analysis was based on the delta-electron density \emph{N}$_{\delta}$ (see Fig.~\ref{fig:1}). The table gives the charge - topology distribution of 227 detected events in which the total charge of relativistic fragments not accompanied by target fragments and product mesons (so - called white stars, \emph{N}$_{ws}$) is $\Sigma$\emph{Z}$_{fr} =$ 6. Such events refer to the most peripheral interactions on Ag and Br nuclei. For the sake of comparison, the distribution of 627 events of $^{10}$C fragmentation that are accompanied by target fragments, \emph{N}$_{tf}$, is given in the table.\par
\begin{figure*}[!ht]
\tiny
\includegraphics[width=4in]{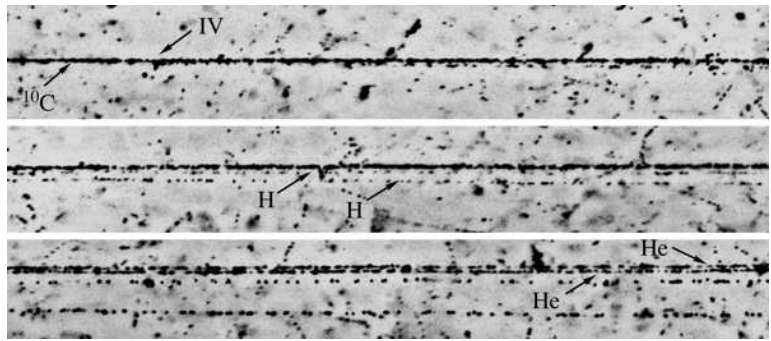}
\caption{Successive macrophotographs of an event involving the dissociation of an $^{10}$C nucleus at an energy of 1.2 GeV per nucleon. The arrows indicate the track of a beam $^{10}$C nucleus, an interaction vertex (IV; at the top), and tracks of H and He fragments.}
\label{fig:2}
\end{figure*}

\indent Figure \ref{fig:2} shows a macrophotograph of one white star. The interaction vertex at which a group of fragments is produced is indicated in the upper photograph. As we move downward, two H fragments and two He fragments become distinguishable in, respectively, the middle and lower photograph. The H track that has largest angle of deflection from the beam-nucleus track arose in the dissociation process $^{10}$C $\rightarrow$ $^9$B$_{g.s.} +$ \emph{p}. The remaining tracks correspond to the decay of $^9$B$_{g.s.}$ unbound nuclei. The pair of He tracks correspond to the decay of another unbound nucleus, $^8$Be$_{g.s.}$.\par
\indent As might have been expected for the isotope $^{10}$C, the dominance of the 2He $+$ 2H channel is the main special feature of the distribution of $\Sigma$\emph{Z}$_{fr} =$ 6 white stars. The branching fraction of this channel is 82\%. Channels characterized by higher thresholds have substantially smaller branching fractions. This picture undergoes a substantial change for events accompanied by target fragments, \emph{N}$_{tf}$.\par
\begin{figure}[!ht]
\begin{center}
\includegraphics[width=70mm]{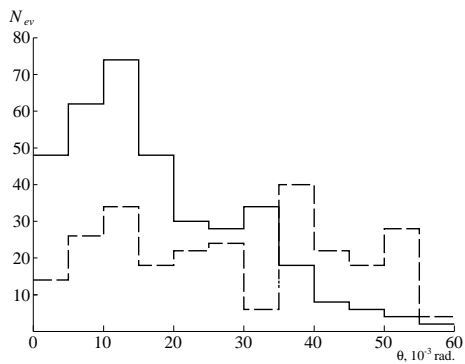}
\caption{Distribution of fragments in white stars with respect to the polar emission angle $\theta$ for the channel $^{10}$C $\rightarrow$ 2He $+$ 2H (dashed-line histogram for H and solid-line histogram for He).}
\label{fig:3}
\end{center}
\end{figure}

\indent Angular measurements of tracks were performed for 184 2He $+$ 2H white stars. Figure \ref{fig:3} shows the distributions of polar emission angles \emph{$\theta$} for H and He fragments. The parameters that describe their Rayleigh distributions are \emph{$\sigma_{\theta H}$} $=$ (51 $\pm$ 3) $\times$ 10$^{-3}$ rad and \emph{$\sigma_{\theta He}$} $=$ (17 $\pm$ 1) $\times$ 10$^{-3}$ rad. These values agree with the values predicted by the statistical model \cite {08,09}, which are \emph{$\sigma_{\theta p}$} $\approx$ 47 $\times$ 10$^{-3}$ rad and \emph{$\sigma_{\theta \alpha}$} $\approx$ 19 $\times$ 10$^{-3}$ rad for, respectively, $^1$H and $^4$He fragments.\par
\begin{figure}[!ht]
\begin{center}
\includegraphics[width=70mm]{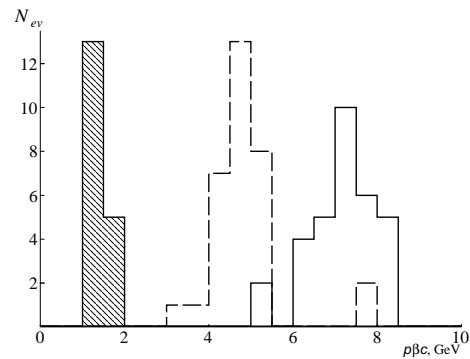}
\caption{ Distribution of \emph{p$\theta$c} for fragments from the $^{10}$C $\rightarrow$ 2He $+$ 2H white stars (solid-line histogram for He and shaded-line histogram for H) and the
$^9$C $\rightarrow$ $^3$He white stars (dashed-line histogram).}
\label{fig:4}
\end{center}
\end{figure}

\indent In order to separate the relativistic isotopes of H and He in the mass number \emph{A}$_{fr}$, one employs, in track-emulsion experiments, measurements of the product of the total momentum of a particle and its velocity, \emph{p$\beta$c}. This product is estimated on the basis of the average angle of multiple Coulomb scattering. 
The relative error in determining \emph{p$\beta$c} is about 20 to 30\%, which is commensurate with the relative mass difference between the $^3$He nucleus and the $^4$He nucleus (alpha particle). In determining \emph{p$\beta$c}, it is necessary to employ tracks of length 2 to 5 cm. This condition prevents the use of the entire available sample of interactions. An identification of the isotopic composition of H and He fragments was performed for 16 2He $+$ 2H white stars (see Fig.~\ref{fig:4}). Also shown for the sake of comparison is the distribution of \emph{p$\beta$c} values measured for $^3$He fragments from events of the fragmentation process $^9$C $\rightarrow$ 3$^3$He \cite {04}. The separation of $^3$He and $^4$He fragments on the basis of measured \emph{p$\beta$c} values is quite unambiguous. Thus, the assumption that, in the sample of 2He $+$ 2H white stars, He and H nuclei are those of the isotopes $^4$He and $^1$H, respectively, is justified. By and large, the charge topology of $\Sigma$~\emph{Z}$_{fr}=$~6 white stars and the dominance of the isotopes $^1$H and $^4$He in them confirms that the formation of the beam of $^{10}$C nuclei was correct. Therefore all of the observed $\Sigma$ \emph{Z}$_{fr} =$ 6 white stars were associated precisely with the dissociation of $^{10}$C nuclei.\par
\begin{figure}[!ht]
\includegraphics[width=70mm]{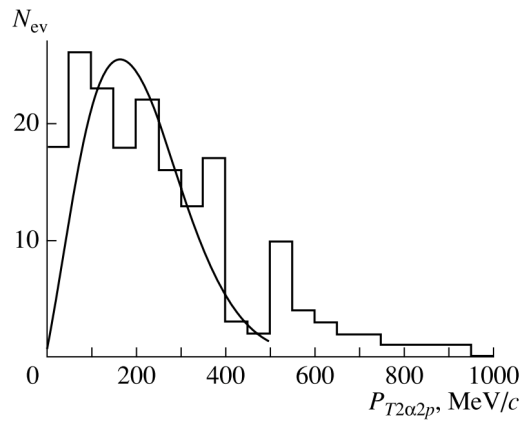}
\caption{Distribution of events of the channel $^{10}$C $\rightarrow$ 2$\alpha$ $+$ 2\emph{p} with respect to the total transverse momentum \emph{P$_{T2\alpha2p}$}. The curve represents the respective Rayleigh distribution.}
\label{fig:5}
\end{figure}

\indent Measurements of angles of relativistic fragments make it possible to estimate the transverse momenta of these fragments according to the expression \emph{P$_{T}$ $\approx$ A$_{fr}P_{0}$}\emph{$sin~\theta$}, where \emph{P}$_{0}$ is the primary momentum per nucleon, which is 2 GeV$/c$ \emph{P$_{T 2\alpha2p}$} per nucleon. The vector sums of the components of transverse momenta yield values of the total momentum transfer. The distribution of these  events with respect to the total transverse momentum \emph{P$_{T}$} (Fig.~\ref{fig:5}) is described by the Rayleigh distribution whose parameter has a value of \emph{$\sigma_{PT}$}(2$\alpha +$2\emph{p}) $=$ 161 $\pm$ 13 MeV$/c$, which is characteristic diffractive dissociation \cite{10}.\par
\begin{figure*}[!ht]
\tiny
\includegraphics[width=5.5in]{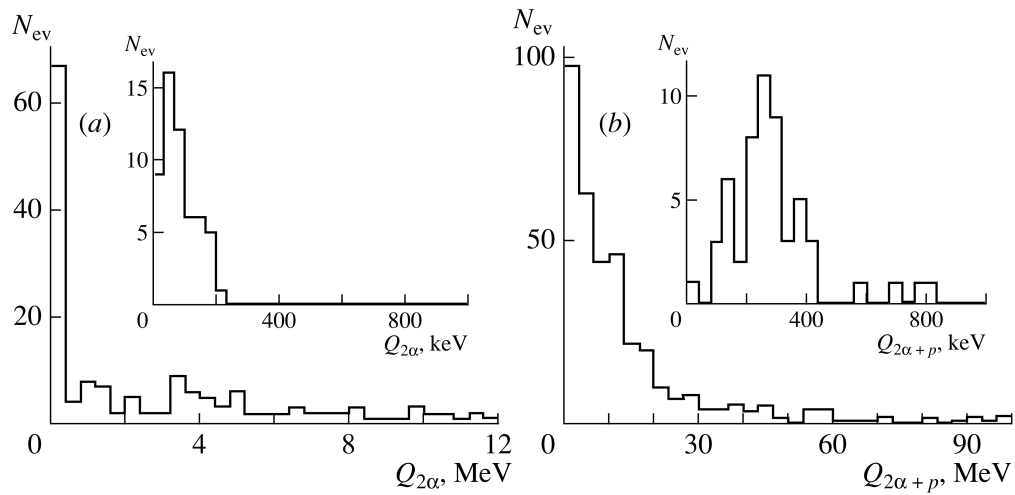}
\caption{Distributions of $^{10}$C $\rightarrow$ 2$\alpha$ $+$ 2\emph{p} events with respect to the (\emph{a}) energy \emph{Q}$_{2\alpha}$ of alpha-particle pairs and (\emph{b}) energy \emph{Q}$_{2\alpha p}$ of the $2\alpha$ $+$ \emph{p} three-particle systems. The insets show enlarged distributions of \emph{Q$_{2\alpha}$} and \emph{Q$_{2\alpha p}$}.}
\label{fig:6}
\end{figure*}

\begin{table}[!ht]
\caption{Charge-topology distribution of fragments from white stars, \emph{N}$_{ws}$ , where the total charge of relativistic fragments is $\Sigma$\emph{Z}$_{fr}$ $=$ 6, and from $\Sigma$\emph{Z}$_{fr}$ $=$ 6 events, \emph{N}$_{tf}$ , accompanied by target fragments or product mesons}
\begin{center}
\begin{tabular}{l|c|c} \hline
~\ Channel ~\ & ~\ \emph{N}$_{ws}$,\% ~\ & ~\ \emph{N}$_{tf}$,\% ~\ \\ \hline
~\ 2He $+$ 2H ~\ & ~\ 186~(81.9) ~\  & ~\ 361~(57.6) ~\ \\ 
~\ He $+$ 4H ~\ & ~\ 12~(5.3) ~\ & ~\ 160~(25.5) ~\ \\
~\ 3He ~\ & ~\ 12~(5.3) ~\ & ~\ 15~(2.4) ~\ \\
~\ 6H ~\ & ~\ 9~(4.0) ~\ & ~\ 30~(4.8) ~\ \\
~\ Be $+$ He ~\ & ~\ 6~(2.6) ~\ & ~\ 17~(2.7) ~\ \\
~\ B $+$ H ~\ & ~\ 1~(0.4) ~\ & ~\ 12~(1.9) ~\ \\
~\ Li $+$ 3H ~\ & ~\ 1~(0.4) ~\ & ~\ 2~(0.3) ~\ \\
~\ $^9$C $+$ \emph{n} ~\ & ~\ \--- ~\ & ~\ 30 (4.8) ~\ \\
\hline
\end{tabular}
\end{center}
\end{table}
\indent The excitation energy of the system of fragments is defined as the difference of the invariant mass of the fragmenting system and the primary-nucleus mass, \emph{Q $=$ M$^{*} -$ M}. The invariant mass of the system of fragments, \emph{M}$^*$, momentum transfer is determined on the basis of the scalar product \emph{M}$^{*2} =$ ($\Sigma$ \emph{P}$_{j}$)$^{2} =$ $\Sigma$ (\emph{P}$_{i} \cdot$ \emph{P}$_{k}$), where \emph{P$_{i,k}$} are the fragment 4-momenta defined in the approximation of conservation of the momentum per nucleon of the parent nucleus. Figure \ref{fig:6} shows the distributions of events of the channel $^{10}$C $\rightarrow$ 2$\alpha +$ 2\emph{p} with respect to the excitation energy \emph{Q$_{2\alpha}$} of alpha-particle pairs and with respect to the excitation energy \emph{Q}$_{2\alpha p}$ of the \emph{$2\alpha$p} three-particle system. Earlier, an analysis of the spectra of \emph{Q$_{2\alpha}$} in the fragmentation of relativistic $^9$Be nuclei permitted reliably revealing the formation of unbound $^9$Be nuclei in the ground state and in the first excited state \cite{11,12}.\par
\begin{figure}[!ht]
\begin{center}
\includegraphics[width=60mm]{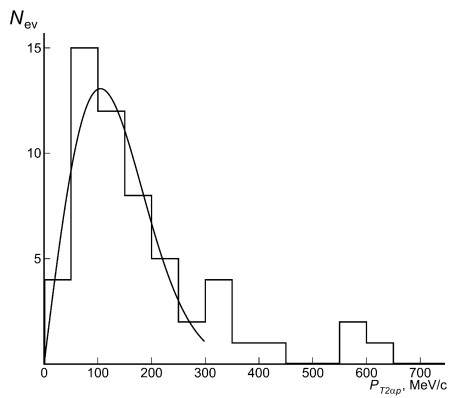}
\caption{Distribution of the total transverse momentum of the \emph{2$\alpha~p$} three-particle system, \emph{P$_{T2\alpha~p}$} , in 2$\alpha$ $+$ 2\emph{p} events involving the formation of $^9$B nuclei. The curve represents the results of the calculations based on the statistical model.}
\label{fig:7}
\end{center}
\end{figure}
\indent In just the same way as in the case of the reaction $^9$Be $\rightarrow ^8$Be$_{g.s.}$, pairs of alpha particles that have the divergence angle not exceeding $10^{-2}$ rad are observed for 68 $^{10}$C $\rightarrow$ 2$\alpha$ $+$ 2\emph{p} white stars. The distribution of \emph{Q$_{2\alpha}$} (Fig.~6a) gives sufficient grounds to conclude that these events lead to the production of $^8$Be nuclei in the ground state, $^8$Be$^{g.s.}$. This is confirmed by mean value of $<$\emph{Q$_{2\alpha}$}$> =$ 63 $\pm$ 30 keV in them, the respective root-mean-square value being 83 keV (see the inset in Fig.~6a). In turn, the distribution of \emph{Q$_{2\alpha p}$} (Fig.~6b) indicates that the dissociation process $^{10}$C $\rightarrow$ 2$\alpha$ $+$ 2\emph{p} is accompanied by the production of an unbound $^9$B nucleus in the ground state, $^9$B$_{g.s.}$. The mean value of $<$\emph{Q$_{2\alpha p}$}$> =$ 254 $\pm$ 18 keV and the root-mean-square value of 96 keV (see the inset in Fig.~6b) are close to the energy of the decay $^9$B$_{g.s.} \rightarrow ^8$Be$_{g.s.} +$ \emph{p} and the respective decay width. There is nearly perfect compliance in the emergence of $^8$Be$_{g.s.}$ (\emph{Q$_{2\alpha}$} $<$ 250 keV) and $^9$B$_{g.s.}$ (\emph{Q$_{2\alpha p}$} $<$ 500 keV), and this is indicative of a cascade character of the process $^{10}$C $\rightarrow$ $^9$B$_{g.s.} \rightarrow$ $^8$Be$_{g.s.}$. The fraction of such events in the sample of $^{10}$C $\rightarrow$ 2$\alpha$ $+$ 2\emph{p} white stars was (30 $\pm$ 4)\%. We can conclude that, in the luster structure of the $^{10}$C nucleus, the unstable nucleus of $^9$B manifests itself with a probability of about 25\%.\par
\begin{figure}[!ht]
\begin{center}
\includegraphics[width=60mm]{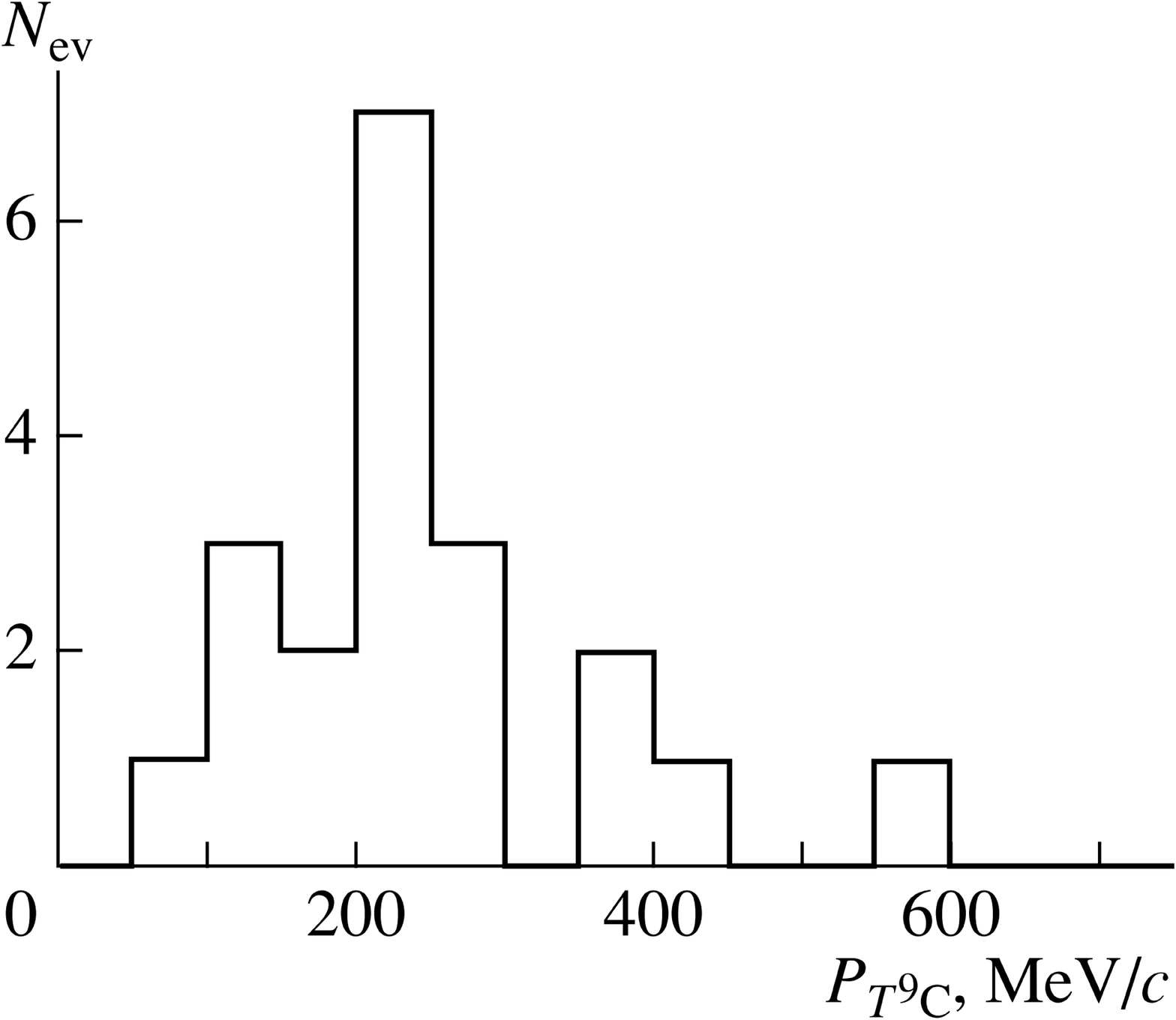}
\caption{Distribution of the transverse momentum \emph{P$_{T^{9}C}$} of $^9$C nuclei in the fragmentation reaction $^{10}$C $\rightarrow$ $^9$C.}
\label{fig:8}
\end{center}
\end{figure}
\indent The distribution of the total transverse momentum \emph{P$_{T2\alpha p}$} of the 2\emph{$\alpha$ p} three-particle system from $^{10}C$ $\rightarrow$ $^9$B white stars (Fig.~\ref{fig:7}) furnishes an argument in favor of a manifestation of the $^9$B nucleus as a component of the structure of the $^{10}$C nucleus. For a group of 40 events (73\%), \emph{$\sigma_{PT^{9}B}$} is 92 $\pm$ 15 MeV$/c$, which corresponds to the value expected on the basis of the statistical model (93 MeV$/c$) \cite{08,09}. Within this model, the radius of the region from which the outer proton is emitted by the $^{10}$C nucleus is \emph{R$_{p}$} $=$ 2.3 $\pm$ 0.4 fm, which is compatible with the value extracted from the measured inelastic cross section on the basis of the geometric-overlap model \cite{13}.\par
\indent The above estimates of \emph{$\sigma_{PT^{9}B}$} and \emph{R$_{p}$} can be compared with data on the fragmentation of $^{10}$C nuclei to $^9$C. As such events, we classify interactions that lead to the formation of target fragments and mesons and in which a heavy fragment retains the charge of the primary nucleus (see table). In 21 interaction events of this type, we did not observe more than one b or g particle, and this gives sufficient grounds to class them with the cases of neutron knockout from $^{10}$C nuclei. Figure~\ref{fig:8} shows the distribution of transverse momenta of $^9$C nuclei (\emph{P$_{T^{9}C}$}), which is characterized by \emph{$\sigma_{PT^{9}C}$} $=$ 224 $\pm$ 49 MeV$/c$. Thus, the spectrum of \emph{P$_{T^{9}C}$} for $^9$C nuclei proves to be substantially harder than the spectrum of \emph{P$_{T2\alpha p}$} for $^9$B nuclei. This circumstance is associated with the knockout of neutrons, whose binding energy is substantially higher than that of outer protons. An estimation of the radius of the neutron-knockout region on the basis of the statistical model yields a value of 1.0 $\pm$ 0.2~$fm$. Of course, this model disregards the clustering of nucleons in the $^{10}$C nucleus. Nevertheless, it provides an indication that the spatial distribution of neutrons in the $^{10}$C nucleus is more compact than the distribution of protons.\par
\begin{figure}
\begin{center}
\includegraphics[width=60mm]{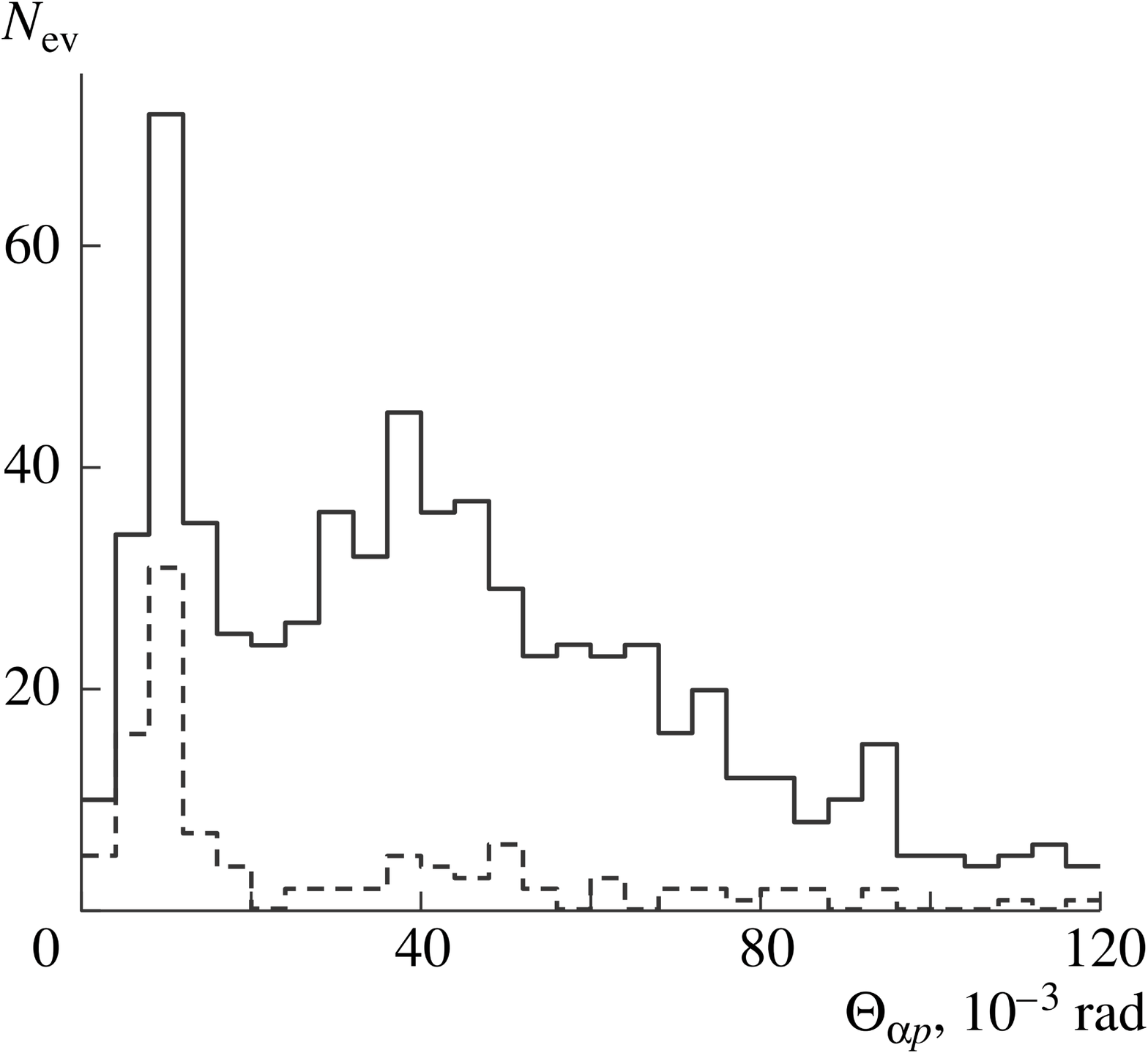}
\caption{Distribution of the opening angle \emph{$\Theta$~$_{\alpha~p}$} between an alpha particle and a proton that arise as fragments (solid-line histogram) and distribution of \emph{$\Theta$~$_{\alpha~p}$} in events involving the production of $^9$B and $^8$Be nuclei (dashed-line histogram).}
\label{fig:9}
\end{center}
\end{figure}
\begin{figure}[!ht]
\includegraphics[width=70mm]{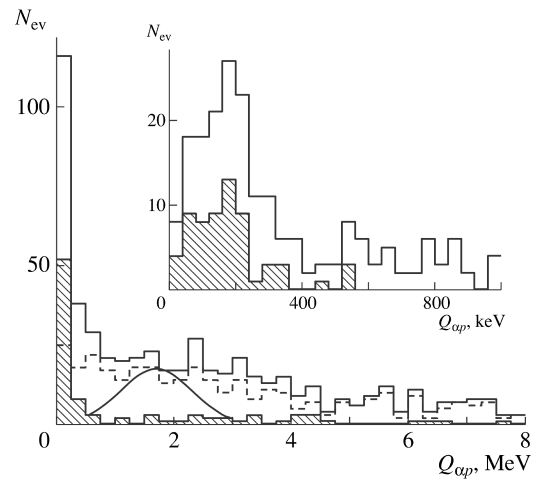}
\caption{Excitation-energy distribution of fragment pairs formed by an alpha particle and a proton in $^{10}$C $\rightarrow$ 2\emph{$\alpha$} $+$ 2\emph{p} white stars: (solid-line histogram) distribution of all \emph{Q$_{\alpha~p}$} combinations, (dashed-line histogram) \emph{Q$_{\alpha~p}$} in events not involving the production of $^9$B and $^8$Be nuclei, (shaded histogram) \emph{Q$_{\alpha~p}$} in events involving the production of $^9$B and $^8$Be nuclei, and (curve) expected position of the $^5$Li resonance. The inset shows an enlarged distribution of \emph{Q$_{\alpha~p}$}.}
\label{fig:10}
\end{figure}

\indent The distribution of the divergence angles \emph{$\Theta_{\alpha p}$} for 736 \emph{$\alpha p$} pairs makes it possible to estimate the contribution to the dissociation of $^{10}$C from the resonance decay $^5$Li$_{g.s.}$ $\rightarrow$ $\alpha$ $+$ \emph{p} (Fig.~\ref{fig:9}). A narrow peak and a broad maximum, which are clarified in the distribution of the excitation energy \emph{Q$_{\alpha p}$} of \emph{$\alpha p$} pairs (Fig.~\ref{fig:10}), are features characteristic of the $^5$Li resonance. The peak owes its existence to the decays of $^9$B nuclei. Further, \emph{$\alpha p$} pairs from the region of 20 $\times$ 10$^{-3}$ $<$ $\Theta$ \emph{$_{\alpha p}$} $<$ 45 $\times$ 10$^{-3}$ rad are grouped in the \emph{Q$_{\alpha p}$} region corresponding to $^5$Li decays. Their distribution is described by a Gaussian functions that is characterized by a mean value of 1.9 $\pm$ 0.1 MeV and by \emph{$\sigma$} $=$ 1.0 MeV. These parameter values comply with the mass (1.7 MeV) and width (1.0 MeV) of the $^5$Li resonance. According to the Gaussian function with the resonance parameters (see Fig.~\ref{fig:10}, approximately 110 \emph{$\alpha p$} pairs could be associated with $^5$Li$_{g.s.}$ decays. The contribution from \emph{Q$_{\alpha p}$} values smaller than the value corresponding to the maximum presumably stemming from the decays of the $^6$Be resonance is present. We were unable to separate a signal of the $^6$Be resonance.\par
\begin{figure}
\includegraphics[width=45mm]{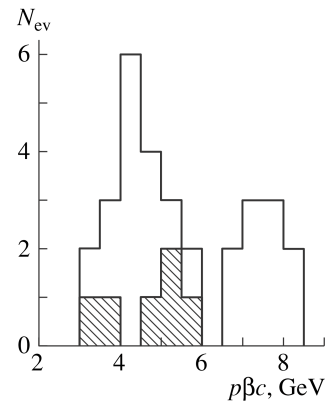}
\caption{Distribution of \emph{p$\beta$c} for He fragments from white stars identified as (solid-line histogram) 2$^3$He $+$ $^4$He and (shaded histogram) $^7$Be $+$ $^3$He events.}
\label{fig:11}
\end{figure}
\indent Among white stars, we observed Be $+$ He and 3He events (see table), which, for the $^{10}$C nucleus, have the thresholds of 15 and 17 MeV. The identification of He fragments on the basis of the parameter \emph{p$\beta$c} (Fig.~\ref{fig:11}) confirms the interpretation of these events as $^7$Be $+$ $^3$He and 2$^3$He $+$ $^4$He and is compatible with the assumption that it is precisely $^{10}$C nuclei that undergo dissociation. The population of these states requires the transition of a neutron from an alpha-particle cluster to the nascent $^3$He cluster. Yet another possibility consists in the presence of deeply bound $^7$Be $+$ $^3$He and 2$^3$He $+$ $^4$He cluster states with a weight of 8\% in the ground state of the $^{10}$C nucleus. The total-transverse-momentum (\emph{P$_{T}$}) distribution of these events is described by the Rayleigh distribution characterized by parameter values of \emph{$\sigma_{PT}$}($^7$Be $+$ $^3$He) $=$ 152 $\pm$ 62 MeV$/c$ and \emph{$\sigma_{PT}$}(2$^3$He $+$ $^4$He) $=$ 204 $\pm$ 65 MeV$/c$.\par
\indent A unique cluster structure of the $^{10}$C nucleus leads to a specific character of its dissociation. In the most peripheral events of the dissociation of $^{10}$C nuclei, about 80\% of events are associated with the channel $^{10}$C $\rightarrow$ 2$\alpha$ $+$ 2\emph{p}. Moreover, it was found that about 30\% of these events refer to the cascade process of the dissociation of $^{10}$C nuclei to $^9$B$_{g.s.}$ $+$ \emph{p}, whereupon the unbound nucleus of $^9$B undergoes a decay to $^8$Be$_{g.s.}$ $+$ \emph{p}. The experimental data obtained in our present study may serve for developing and testing the cluster model of the $^{10}$C nucleus.
\begin{center}
ACKNOWLEDGMENTS
\end{center}
\indent This work was supported in part by the Russian Foundation for Basic Research (project no. 12-02-00067) and by grants from the plenipotentiaries of Bulgaria and Romania at the Joint Institute for Nuclear Research (Dubna).

\indent \par
\indent \emph{Translated by A. Isaakyan}

\begin{thebibliography}{99}
\bibitem{01} The BECQUEREL Project, http://becquerel.jinr.ru/
\bibitem{02} N. G. Peresadko et al., Phys. At. Nucl. \textbf{70}, 1226 (2007); nucl-ex/0605014.
\bibitem{03} R. Stanoeva et al., Phys. At. Nucl. \textbf{72}, 690 (2009); arXiv: 0906.4220 [nucl-ex].
\bibitem{04}  D. O. Krivenkov et al., Phys. At. Nucl. \textbf{73}, 2103 (2010); arXiv: 1104.2439 [nucl-ex].
\bibitem{05} D. A. Artemenkov et al., Few-Body Syst. \textbf{50}, 259 (2011); arXiv: 1105.2374 [nucl-ex].
\bibitem{06} D. A. Artemenkov et al., Int. J. Mod. Phys. E \textbf{20}, 993 (2011); arXiv: 1106.1748 [nucl-ex].
\bibitem{07} R. R. Kattabekov, K. Z. Mamatkulov, D. A. Artemenkov, et al., Phys. At. Nucl. \textbf{73}, 2110 (2010); arXiv: 1104.5320 [nucl-ex].
\bibitem{08} H. Feshbach and K. Huang, Phys. Lett. B \textbf{47}, 300 (1973).
\bibitem{09} A. S. Goldhaber, Phys. Lett. B \textbf{53}, 306 (1974).
\bibitem{10} N. G. Peresad'ko, V. N. Fetisov, Yu. A. Aleksandrov, et al., JETP Lett. \textbf{88}, 75 (2008); arXiv: 1110.2881 [nucl-ex].
\bibitem{11}  D. A. Artemenkov et al., Phys. At. Nucl. \textbf{70}, 1222 (2007); nucl-ex/0605018.
\bibitem{12} D. A. Artemenkov et al., Few-Body Syst. \textbf{44}, 273 (2008).
\bibitem{13} A. Ozawa, T. Suzuki, and I. Tanihata, Nucl. Phys. A \textbf{693}, 32 (2001).
\end{thebibliography}
\end{document}